\documentclass{PoS}

\usepackage{ae}
\usepackage{aecompl}
\usepackage{graphicx}
\usepackage{fancyhdr}
\usepackage{amsmath}

\def\pa{\partial}
\def\nn{\nonumber \\}

\def\hathH{\hat{h}\hat{H}}
\def\haths{\hat{h}\hat{s}}
\def\hatHs{\hat{H}\hat{s}}

\title{Blind spots for neutralinos in NMSSM with light singlet scalar}

\ShortTitle{Blind spots for neutralinos in NMSSM with light singlet scalar}

\author{M. Badziak, M. Olechowski, \speaker{P. Szczerbiak}
         \\
        Institute of Theoretical Physics, Faculty of Physics, University of Warsaw\\
		ul. Pasteura 5, PL--02--093 Warsaw, Poland\\
        E-mail: 
        \email{mbadziak@fuw.edu.pl},
        \email{Marek.Olechowski@fuw.edu.pl},
        \email{Pawel.Szczerbiak@fuw.edu.pl}}

\abstract{
A substantial contribution to the SM-like Higgs mass may come from the mixing with the lightest singlet-dominated scalar in NMSSM for moderate or large values of $\tan\beta$~\cite{BaOlPo}. The LSP neutralino in this model may also contain the singlet component~-- singlino. 
In this work we analyze the direct detection cross section for a Higgsino-singlino LSP with the emphasis on possible direct detection blind spots in the parameter space. Correlations between the LSP properties and the effects in the Higgs sector coming from the mixing are also discussed.
}

\FullConference{18th International Conference From the Planck Scale to the Electroweak Scale \\
		 25-29 May 2015\\
		 Ioannina, Greece }

\begin{document}

\section{Introduction \label{introduction}}

In recent years the Next-to-Minimal Supersymmetric Standard Model (NMSSM) gained a lot of attention in particle physics community. The primary reason for this is the
measurement of the Higgs mass of 125 GeV which is hard to accommodate in the Minimal Supersymmetric Standard Model (MSSM) with light stops, hence threatening
naturalness of supersymmetry. On the other hand, in NMSSM the 125 GeV Higgs can be compatible with relatively light stops due to additional contributions to the
Higgs mass. The Higgs mass in NMSSM is usually enhanced by demanding large singlet-Higgs-Higgs superpotential coupling $\lambda$ and small $\tan\beta$.
However, it was recently emphasized that light stops in NMSSM are also possible for moderate and large $\tan\beta$
provided that the mixing between the Higgs and a lighter singlet-dominated scalar is non-negligible \cite{BaOlPo}.

Presence of a light singlet-dominated scalar has important implications for neutralino dark matter. This is because the Higgs-singlet mixing
modifies the Higgs couplings to LSP and nucleons, hence also the spin-independent (SI) scattering cross-section $\sigma_{\rm SI}$. 
Moreover, a light singlet-dominated scalar can mediate the LSP-nucleon interaction itself giving contribution to $\sigma_{\rm SI}$ that can be even
larger than the one coming from the Higgs exchange. In these proceedings we investigate the SI scattering cross-section and derive
analytic formulae for the neutralino blind spots i.e. regions of the NMSSM parameter space with strongly suppressed SI scattering cross-section, including
the effects of the singlet. We focus on a singlino-Higgsino LSP since this is the most distinct from MSSM type of LSP and, in addition, can be a thermal relic
with the abundance in agreement with the observations. Many results presented in these proceedings (except subsection~\ref{model_mup=0}) are discussed in more details in~\cite{BS_NMSSM}.

\section{CP-even scalar and neutralino sector of the NMSSM \label{NMSSM_sectors}}

We will start from a~brief summary of the NMSSM CP-even scalar and neutralino sector. Let us parametrize the NMSSM specific part of the superpotential as:
\begin{equation}
\label{W}
W_{\rm NMSSM}= \lambda SH_u\cdot H_d + f(S) \,,
\end{equation}
where $S$ is a SM-singlet superfield. In a general model the function $f(S)$ contains all possible renormalizable terms
$f(S)=\xi_F S +\mu' S^2/2 + \kappa S^3/3$.
The corresponding soft terms are:\footnote{ 
We use the same notation for chiral superfields as for 
their scalar components; $H_u\cdot H_d\equiv H_u^+H_d^--H_u^0H_d^0$.
}  
\begin{align}
-{\cal{L}}_{\rm soft}
\supset
&\,\,
m_{H_u}^2H_u^\dagger H_u +m_{H_d}^2H_d^\dagger H_d+m_{S}^2\left|S\right|^2
\nn
&\label{Lsoft}
+\left(
A_\lambda\lambda H_u\cdot H_d S +\frac13A_\kappa\kappa S^3
+m_3^2H_u\cdot H_d + \frac12 m_S'^2S^2 + \xi_SS
+{\rm h.c.}\right)\,.
\end{align}
The first term in (\ref{W}) is the source of the effective Higgsino 
mass parameter: $\mu\equiv\lambda v_s$ (note that we dropped the usual MSSM $\mu$ parameter by a~shift symmetry). 
In the simplest version, known as the scale-invariant NMSSM, 
$m_3^2=m_s'^2=\xi_S=0$ while $f(S)\equiv\kappa S^3/3$.
 
There are three neutral CP-even scalar fields, 
$H_u^0$, $H_d^0$, $S$ which are the real parts of excitations 
around the real vevs, $v_u\equiv v \sin\beta$, $v_d\equiv v \cos\beta$, 
$v_s$ with $v^2=v_u^2 + v_d^2\approx (174 {\rm GeV})^2$,  
of the neutral components of doublets $H_u$, $H_d$ and the singlet 
$S$.
It is more convenient for us to work in the basis 
$\left(\hat{h}, \hat{H}, \hat{s}\right)$, where 
$\hat{h}=H_d^0\cos\beta + H_u^0\sin\beta$, 
$\hat{H}=H_d^0\sin\beta - H_u^0\cos\beta$ and $\hat{s}=S$. 
The $\hat{h}$ field has exactly the same couplings to the gauge bosons 
and fermions as the SM Higgs field. In this basis the off-diagonal elements of the scalar mass squared matrix~$M^2$ have the form:
\begin{equation}
\label{tilde_M^2_diag}
{M}^2_{\hathH} = \frac{1}{2}(M^2_Z-\lambda^2 v^2)\sin4\beta, \quad
{M}^2_{\haths} =  \lambda v (2\mu-\Lambda \sin2\beta), \quad
{M}^2_{\hatHs} = \lambda v \Lambda \cos2\beta ,
\end{equation}
and $\Lambda \equiv A_{\lambda}+\langle\pa_S^2 f\rangle$. 
The mass eigenstates of ${M}^2$, denoted by $h_i$ (with $h_i=h,H,s$), 
are expressed in terms of the hatted fields with the help of the 
diagonalization matrix $\tilde{S}$:\footnote
{
The matrix $\tilde{S}$ is related to the commonly used Higgs mixing 
matrix $S$ by a rotation by the angle $\beta$ in the 2-dimensional 
space of the weak doublets. 
}
\begin{equation}
\label{hat-S}
h_i
=\tilde{S}_{h_i\hat{h}}\hat{h}
+\tilde{S}_{h_i\hat{H}}\hat{H}
+\tilde{S}_{h_i\hat{s}}\hat{s}
\,.
\end{equation}

The neutralino mass matrix in NMSSM is 5-dimensional.
However, in this work we assume that gauginos are heavy and thus 
we focus on the sub-matrix describing the three lightest neutralinos:
\begin{equation}
\label{M_chi}
 {M_{\chi^0}}=
\left(
\begin{array}{ccc}
  0 & -\mu & -\lambda v \sin\beta \\[4pt]
  -\mu & 0 & -\lambda v \cos\beta \\[4pt]
  -\lambda v \sin\beta & -\lambda v \cos\beta 
& \langle\partial_S^2f\rangle \\
\end{array}
\right) \,.
\end{equation}
Trading the model dependent term $\langle\partial_S^2f\rangle$
for one of the eigenvalues, $m_{\chi_j^0}$, of the above neutralino mass
matrix we find the following (exact at the tree level) relations 
for the neutralino diagonalization matrix elements:
\begin{equation}
\label{Nj3Nj5}
\frac{N_{j3}}{N_{j5}}
=
\frac{\lambda v}{\mu}
\,
\frac{(m_{\chi_j^0}/\mu)\sin\beta-\cos\beta}
{1-\left(m_{\chi_j^0}/\mu\right)^2}
\,,\qquad
\frac{N_{j4}}{N_{j5}}
=
\frac{\lambda v}{\mu}
\,
\frac{(m_{\chi_j^0}/\mu)\cos\beta-\sin\beta}
{1-\left(m_{\chi_j^0}/\mu\right)^2}
\,,
\end{equation}
where $j=1,2,3$ and $|m_{\chi_1^0}|\le|m_{\chi_2^0}|\le|m_{\chi_3^0}|$.
Notice that the physical (positive) LSP mass is equal 
$m_{\rm LSP}\equiv|m_{\chi}|$
(to simplify the notation from now on we use $\chi\equiv\chi_1^0$, 
$m_{\chi}\equiv m_{\chi_1^0}$ etc.).

In our discussion we will consider only positive values of $\lambda$. 
The results for negative $\lambda$ are exactly the same due to the invariance 
under the transformation $\lambda\to-\lambda$, 
$\kappa\to-\kappa$, $\xi_S\to-\xi_S$, $\xi_F\to-\xi_F$, $S\to-S$ 
with other fields and couplings unchanged.

\section{Spin-independent scattering cross-section \label{SI_cross_section}}

The spin-independent cross-section for the LSP interacting 
with the nucleus with the atomic number $Z$ and the mass number $A$ 
is given by
\begin{equation}
\sigma_{\rm SI}
=
\frac{4\mu^2_{\rm red}}{\pi}\,\frac{\left[Zf^{(p)}+(A-Z)f^{(n)}\right]^2}{A^2}
\,,
\end{equation}
where $\mu^2_{\rm red}$ is the reduced mass of the nucleus and the LSP. Usually, the experimental limits concern the cross section $\sigma_{\rm SI}$ defined as 
$\frac12(\sigma_{\rm SI}^{(p)}+\sigma_{\rm SI}^{(n)})$. Thus, in the rest of the paper we will follow this convention. When the squarks are heavy the effective couplings $f^{(N)}$ ($N=p,n$) 
are dominated by the t-channel exchange of the CP-even scalars
\cite{JuKaGr}:
\begin{equation}
\label{fN}
f^{(N)}
\approx
\sum_{i=1}^3
f^{(N)}_{h_i}
\equiv
\sum_{i=1}^3
\frac{\alpha_{h_i\chi\chi}\alpha_{h_iNN}}{2m_{h_i}^2}
\,.
\end{equation}
The couplings of the $i$-th scalar to the LSP and to the 
nucleon in our rotated basis ($\hat{h}$,$\hat{H}$,$\hat{s}$) are given, respectively, by
\begin{align}
\alpha_{h_i\chi\chi}
\approx
\sqrt{2}\lambda 
&
\left[
\tilde{S}_{i\hat{h}}N_{15}\left(N_{13}\sin\beta+N_{14}\cos\beta\right)
+\tilde{S}_{i\hat{H}}N_{15}\left(N_{14}\sin\beta-N_{13}\cos\beta\right)
\right.\nn&\,\,\,\,\left.
+\tilde{S}_{i\hat{s}}
\left({N_{13}}{N_{14}}-\frac{\kappa}{\lambda}N_{15}^2\right)
\right]
\,,
\label{alpha-h00_G}
\end{align}
\begin{equation}
\alpha_{h_iNN}
\approx
\frac{m_N}{\sqrt{2}v} 
\left[
\tilde{S}_{i\hat{h}}\left(F^{(N)}_d+F^{(N)}_u\right)
+\tilde{S}_{i\hat{H}} \left(\tan\beta F^{(N)}_d-\frac{1}{\tan\beta} F^{(N)}_u \right)
\right]\,,
\label{alpha-hNN_G}
\end{equation}
where we neglected the contributions from (heavy) gauginos i.e.\ $N_{11}\approx 0\approx N_{12}$ and introduced the form factors  $F^{(p)}_{u}\approx 0.152$, $F^{(p)}_{d}\approx 0.132$, 
$F^{(n)}_{u}\approx 0.147$, $F^{(n)}_{d}\approx 0.140$~\cite{Belanger:2013oya}.

\section{Blind spots coming from the $h$ and $H$ exchange \label{bs_hH}}

In this section we will present the blind spot conditions analogous to 
those already known in the MSSM i.e.\ for decoupled $s$ with $m_s\to\infty$.
Thus, we can neglect the mixing of $h$ and $H$ with $s$ in $f_h^{(N)}$ 
and $f_H^{(N)}$ as well as the $f_s^{(N)}$ amplitude. 
Let us start from the situation in which $f_h^{(N)}$ interferes destructively with the contribution $f_H^{(N)}$ mediated by the heavy 
Higgs doublet. The coupling of $H$ to down quarks, hence also to nucleons, may be enhanced by large $\tan\beta$ which could compensate the suppression of $f_H^{(N)}$ by $m_H^{-2}$. Considering $\tan\beta\gg 1$ we can neglect $\tilde{S}_{h\hat{H}}\sim-2(M_z^2-\lambda^2v^2)/(\tan\beta\,m_H^2)$ (see. eq.~\eqref{tilde_M^2_diag}), which causes $\tilde{S}_{h\hat{h}}\approx\tilde{S}_{H\hat{H}}\approx1$, and rewrite the amplitudes as: 
\begin{equation}
f_h^{(N)}+f_H^{(N)}
\approx
\frac{\lambda m_N N_{15}}{vm_h^2}F^{(N)}
\left[N_{13}\sin\beta+N_{14}\cos\beta
+\left(\frac{m_h}{m_H}\right)^2\frac{\tan\beta}{2}
(N_{14}\sin\beta-N_{13}\cos\beta)\right],
\end{equation}
where we used the approximation $F_d^{(N)}=F_u^{(N)}(\equiv F^{(N)})$.
We can see immediately that if the LSP is a pure singlino or a pure Higgsino 
this expression vanishes. For a~mixed LSP the sum of the above amplitudes may be small only if there is a~cancellation in the square bracket. Perfect cancellation gives the following blind spot condition:
\begin{equation}
\label{bs_fhH_0}
\frac{m_{\chi}}{\mu}-\sin2\beta
\approx \left(\frac{m_h}{m_H}\right)^2\frac{\tan\beta}{2}  
\,.
\end{equation}
This is a similar result to the one obtained in the MSSM~\cite{Wagner}, but for the Higgsino-singlino LSP rather than the Higgsino-gaugino one. Note that $m_\chi\mu>0$ is required in contrast to MSSM. This kind of a blind spot allows for a highly-mixed singlino-Higgsino LSP and large $\tan\beta$ since 
\eqref{bs_fhH_0}
can be satisfied with $|m_{\chi}|\approx |\mu|$ provided that $\left({m_h}/{m_H}\right)^2\tan\beta\sim \mathcal{O}(1)$. In the decoupling limit of $H$, i.e.\ $m_H\to\infty$, our blind spot condition simplifies to:
\begin{equation}
\label{bs_fh_0}
\frac{m_{\chi}}{\mu}-\sin2\beta=0\,.
\end{equation}
This result is analogous to the one obtained in~\cite{Hall} for the 
Higgsino-gaugino LSP in the MSSM (as called traditional blind spots) but, again, with opposite sign between 
the two terms in the l.h.s. Notice that if $\tan\beta$ is not small, 
the blind spot condition implies a singlino-dominated LSP 
($|m_{\chi}|\ll|\mu|$) for which $f_h$ is suppressed anyway.

\subsection{Mixing with the singlet}

Now we can ask the question what happens if $h$ and $H$ mix with the singlet-dominated scalar $s$ (but still in the limit of large $m_s$ i.e.\ with $f_s$ neglected). Let us start with the blind spot condition~\eqref{bs_fh_0}. The effect of $h$-$s$ mixing modifies the r.h.s.\ of that formula giving: 
\begin{equation}
\label{bs_fh_mix}
\frac{m_{\chi}}{\mu}-\sin2\beta
\approx
-\frac{\tilde{S}_{h\hat{s}}}{\tilde{S}_{h\hat{h}}}\,\frac{\mu}{\lambda v}
\left[1-\left(\frac{m_{\chi}}{\mu}\right)^2\right]
\left(\frac{N_{13}}{N_{15}}\frac{N_{14}}{N_{15}}-\frac{\kappa}{\lambda}\right)\,,
\end{equation}
where:
\begin{equation}
\label{Deltamix}
\frac{\tilde{S}_{h\hat{s}}}{\tilde{S}_{h\hat{h}}}
\approx{\rm sgn}(\Lambda\sin2\beta-2\mu)\frac{\sqrt{2|\Delta_{\rm mix}|m_h}}{m_s}\,,\quad\quad
\Delta_{\rm mix} \equiv m_h - \hat{M}_{hh}\,.
\end{equation}
The last parameter, $\Delta_{\rm mix}$, measures the linear correction to the Higgs mass coming from the mixing with other states (mainly $s$). For $m_h<m_s$ this correction is negative which is not preferable. Thus, the requirement of small $|\Delta_{\rm mix}|$, say smaller than $\mathcal{O}(1)$ GeV, implies 
$({\tilde{S}_{h\hat{s}}}/{\tilde{S}_{h\hat{h}}})\lesssim0.1(m_h/m_s)$. 
Therefore, in order to have a strong modification of the blind spot 
condition at least one of 
the factors in the r.h.s.\ of eq.~\eqref{bs_fh_mix} must 
be much larger than 1 which puts limitations on the composition of the LSP. Namely, for a singlino-dominated and highly-mixed singlino-Higgsino LSP the blind spot with $m_\chi\mu < 0$ requires large $|\kappa/\lambda|$, and thus small $\lambda$ if we impose perturbativity limits, while for a Higgsino-dominated LSP such a blind spot can be present both for small and large $\lambda$.  

In the case of a non-negligible effect of $H$ the analogue of eq.~\eqref{bs_fhH_0} takes the form:
\begin{equation}
\label{bs_fhH_mix}
\frac{m_{\chi}}{\mu}-\sin2\beta
\approx
\left(\frac{m_h}{m_H}\right)^2\frac{\tan\beta}{2}
\left[1-\frac{\tilde{S}_{H\hat{s}}}{\tilde{S}_{H\hat{H}}}
\frac{\mu}{\lambda v}
\left(1-\left(\frac{m_{\chi}}{\mu}\right)^2\right)
\left(\frac{N_{13}}{N_{15}}\frac{N_{14}}{N_{15}}-\frac{\kappa}{\lambda}\right)
\right]
\,.
\end{equation}
If $\tilde{S}_{h\hat{s}}$ is not negligible with respect to $\big(\frac{m_h}{m_H}\big)^2\frac{\tan\beta}{2}\;\tilde{S}_{H\hat{s}}$ we should also add here the correction~\eqref{bs_fh_mix}. However, the conclusions for blind spots regarding the LSP composition are qualitatively similar to those resulting from~\eqref{bs_fh_mix} (see~\cite{BS_NMSSM} for detailed analysis).

\section{Blind spots coming from the $h$ and $s$ exchange \label{bs_hs}}

Now we move to the main point of our discussion i.e.\ the situation when the singlet-like scalar is light. To simplify our consideration and feel the qualitative difference with respect to the previous section, we decouple the $H$ state i.e.\ we neglect the $f_H^{(N)}$ amplitude (assuming that $m_s,m_h\ll m_H$) but at the same time we take into account the $\hat{H}$-$\hat{s}$ mixing effects in $f_s^{(N)}$ which may be enhanced by large $\tan\beta$. The modification will concern eqs.~\eqref{bs_fh_0} and \eqref{bs_fh_mix}. In order to shorten our notation we introduce additional parameters:
\begin{equation}
\label{As+cs}
\mathcal{A}_s
\equiv
-\gamma\frac{1+c_s}{1+c_h}\left(\frac{m_h}{m_s}\right)^2
\,,
\qquad\qquad
c_{h_l}\equiv1+\frac{\tilde{S}_{h_l\hat{H}}}{\tilde{S}_{h_l\hat{h}}}\left(\tan\beta-\cot\beta\right)\,,
\end{equation}
where $h_l=h,s$ and 
$\gamma\equiv\tilde{S}_{h\hat{s}}/\tilde{S}_{h\hat{h}}$ is given in~\eqref{Deltamix}. In the limit of large $\tan\beta$ the $c_s$ ($c_h$) parameter measures the ratio of the couplings, normalized to SM values, of the $s$ ($h$) scalar to the $b$ quarks and to the $Z$ bosons.\footnote
{
It is easier to make a light scalar $s$ compatible with the LEP bounds when $|c_s|$ is small \cite{BaOlPo}, especially for $m_s\lesssim85$ GeV. We should note, however, that $c_s<1$ implies $c_h>1$ which in turn leads to suppressed branching ratios of $h$ decaying to gauge bosons, so $c_h$ is constrained by the LHC data.
}
Then our blind spot condition reads:
\begin{equation}
\label{bs_fhs_mix}
\frac{m_{\chi}}{\mu}-\sin2\beta
\approx
-\frac{\gamma+\mathcal{A}_s}{1-\gamma\mathcal{A}_s}\,\frac{\mu}{\lambda v}
\left[1-\left(\frac{m_{\chi}}{\mu}\right)^2\right]
\left(\frac{N_{13}}{N_{15}}\frac{N_{14}}{N_{15}}-\frac{\kappa}{\lambda}\right)\,,
\end{equation}
which is of the form~\eqref{bs_fh_mix} but with a crucial modification of the 
first factor in the r.h.s.:
\begin{equation}
\label{bs_fhs_mix-modif}
\gamma\equiv\frac{\tilde{S}_{h\hat{s}}}{\tilde{S}_{h\hat{h}}}
\quad\longrightarrow\quad
\frac{\gamma+\mathcal{A}_s}{1-\gamma\mathcal{A}_s}\,.
\end{equation}
(Note that for $m_s\to\infty$ we have $\mathcal{A}_s\to 0$.) The new prefactor can be at least one order of magnitude larger than the previous one i.e.\ in the case with only $h$ exchange taken into account, which soften the constraints on the LSP composition e.g.\ large $|\kappa/\lambda|$ is not always necessary in the case of a singlino/highly-mixed singlino-Higgsino LSP and $m_\chi\mu<0$.

\begin{figure}
\center
\includegraphics[width=0.49\textwidth]{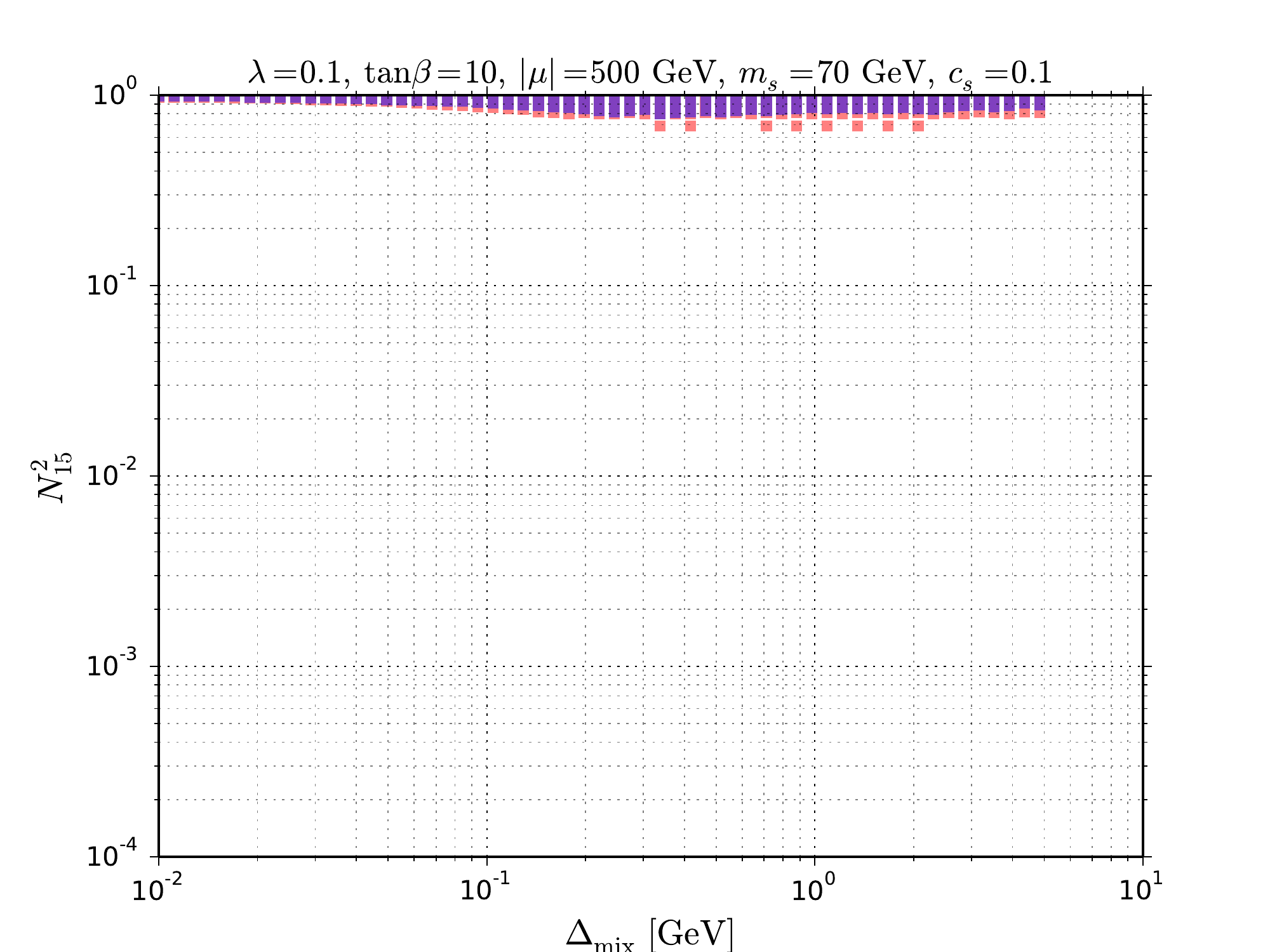}
\includegraphics[width=0.49\textwidth]{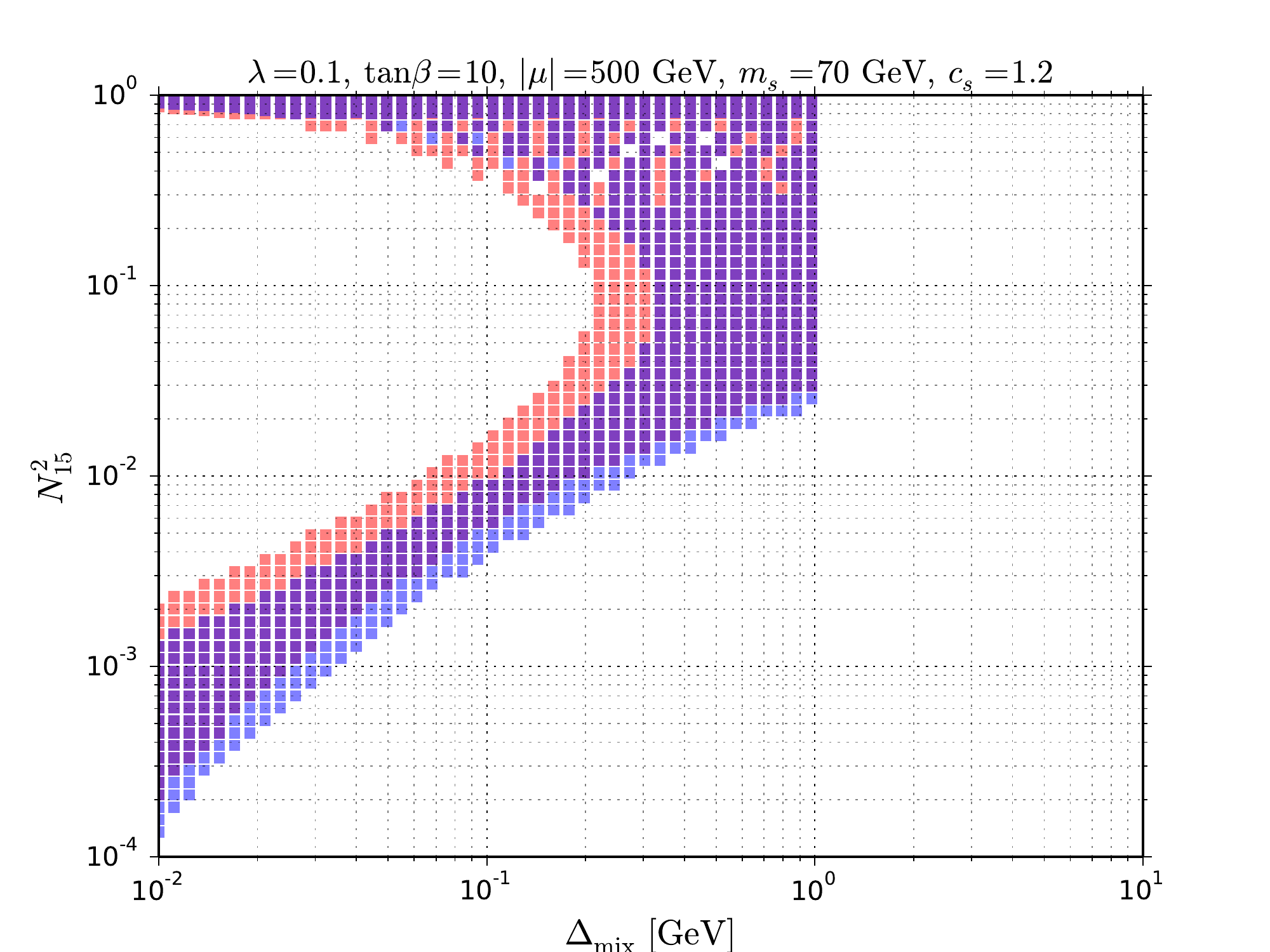}\\
\includegraphics[width=0.49\textwidth]{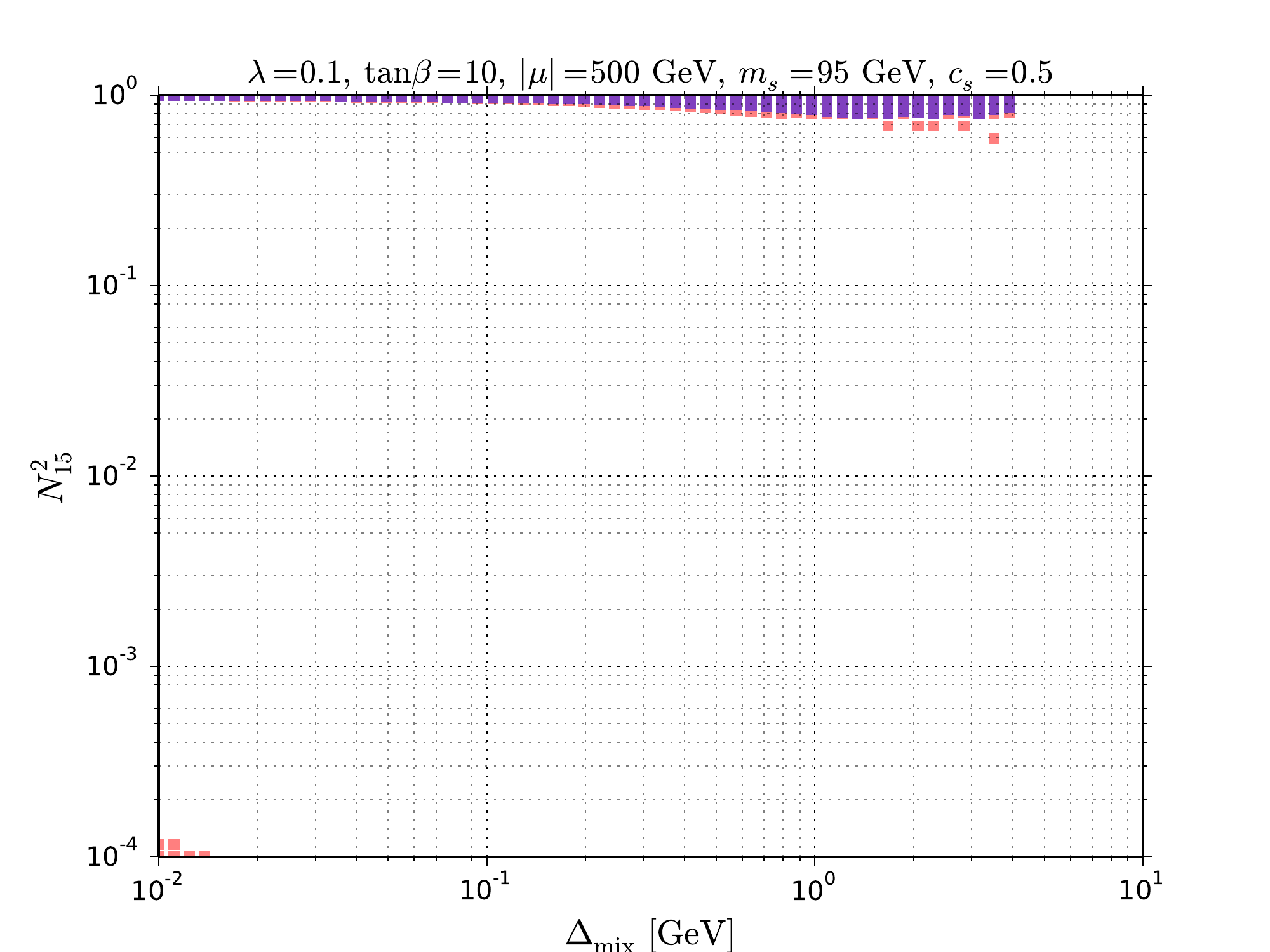}
\includegraphics[width=0.49\textwidth]{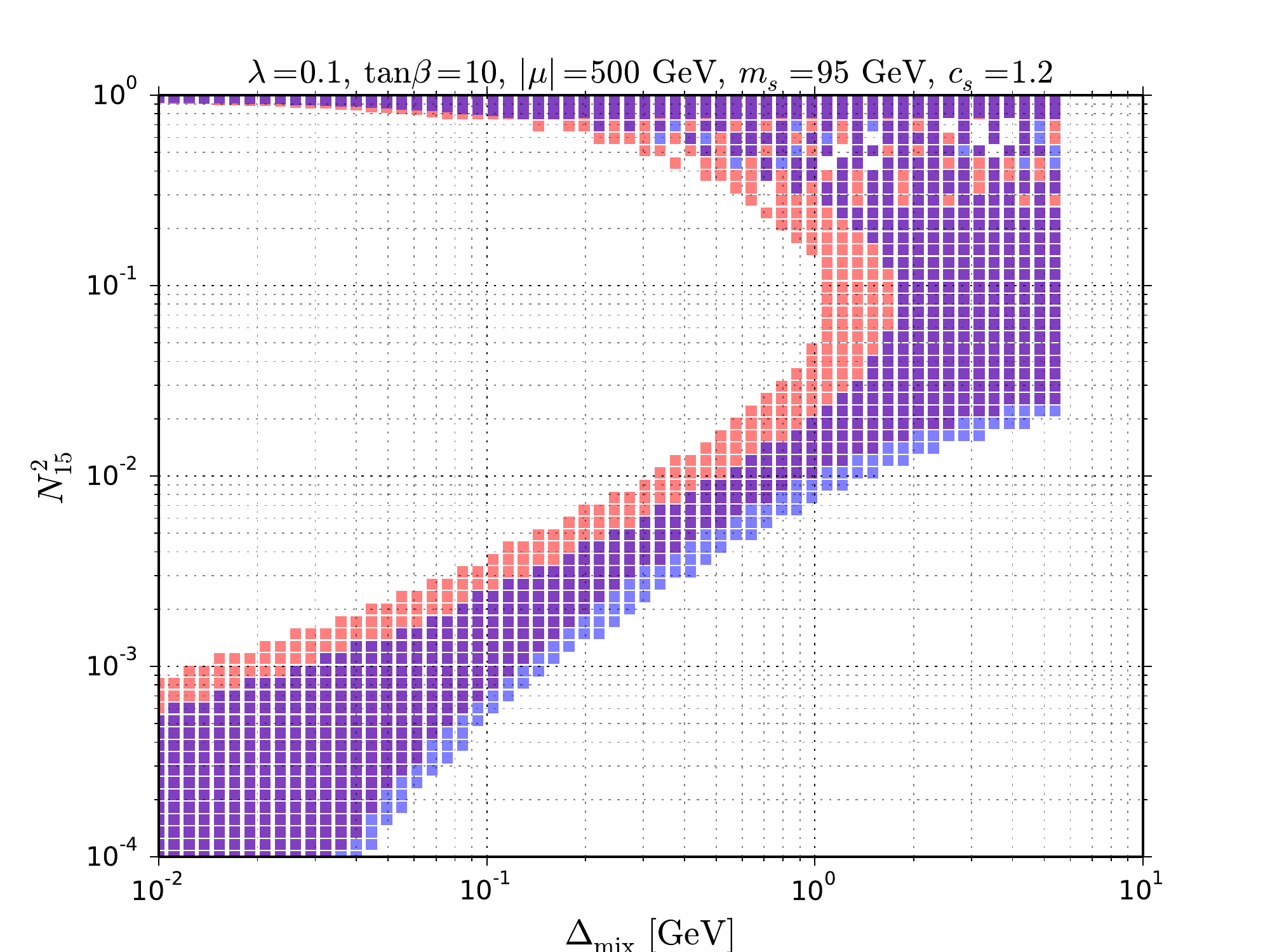}
\caption{Regions of the plane ($\Delta_{\rm mix}-N_{15}^2$) with $\sigma_{\rm SI}$ smaller than the neutrino background~\cite{NeutrinoB} for $m_{\chi}\mu>0$ (red) and
$m_{\chi}\mu<0$ (blue), while keeping $|\kappa|\leq 0.6$. Upper (lower) plots correspond to $m_s=70$ (95) GeV whereas the left (right) to $c_s$ smaller (larger) than 1.
}
\label{fig:bs_fhs_N15_delta}
\end{figure}

From our perspective the most interesting situation takes place when $\Delta_{\rm mix}$, being now positive, is large. As stated above, for $m_s\lesssim 85$~GeV small $|c_s|$ and hence large $\tan\beta$ and small $\lambda$ are preferred~\cite{BaOlPo}. For definiteness, let us consider $\tan\beta=10$, $\lambda=0.1$ and two representative values of $m_s$, 70 and 95 GeV, for which the LEP bounds are, respectively, quite severe and rather mild. In Fig.~\ref{fig:bs_fhs_N15_delta} we present the points (for a few values of $c_s$) for which $\sigma_{\rm SI}$ is smaller than the neutrino background for two signs of $m_\chi\mu$ (in all plots presented in this paper the LEP and LHC Higgs constraints are satisfied, at $2\sigma$ level, unless otherwise stated). The most apparent difference between $c_s>1$ and $c_s<1$ is that in the first case there are no points with Higgsino-dominated LSP, whereas in the second one there is a negative correlation between Higgsino admixture and $\Delta_{\rm mix}$ (for $N_{15}^2\lesssim0.1$). In order to explain this behavior we rewrite the blind spot condition~\eqref{bs_fhs_mix} in the form adequate for the Higgsino-dominated limit i.e.\ for $|m_\chi/\mu|\to 1$. The result reads:
\begin{equation}
\label{bs_fhs_mix-higgsino}
\frac{\gamma+\mathcal{A}_s}{1-\gamma\mathcal{A}_s}\approx
{\rm sgn}(\mu)|N_{15}|\sqrt{2(1-{\rm sgn}\left(m_{\chi}\mu\right)\sin2\beta)}
\,.
\end{equation}
For specific values of $c_s$ and $m_s$ (chosen in our example) the l.h.s. of the above equation is proportional\footnote{It can be easily seen if we notice that $\mathcal{A}_s=-\gamma\;{\rm const}$, where ${\rm const}=\frac{1+c_s}{1+c_h}\big(\frac{m_h}{m_s}\big)^2>1$. Moreover $|\gamma\mathcal{A}_s|\ll 1$ and hence the denominator in the l.h.s. of~\eqref{bs_fhs_mix-higgsino} is roughly 1.}  to $-\gamma$ and thus to $\Delta_{\rm mix}$ (see~\eqref{Deltamix})~-- this explains why there is a correlation between $\Delta_{\rm mix}$ and $|N_{15}|$. To understand why for $c_s>1$ $(c_s<1)$ there are (no) points which fulfill~\eqref{bs_fhs_mix-higgsino} we should notice (see eq.~\eqref{tilde_M^2_diag}) that for $\tan\beta\gg1$ we have ${\rm sgn}(1-c_s)={\rm sgn}(\Lambda\gamma)={\rm sgn}(\mu\gamma)$~-- the second equality holds because a partial cancellation between the two terms in $M_{h\hat s}^2$ is needed.\footnote{
This happens in our example in Fig.~\ref{fig:bs_fhs_N15_delta} because for $|\mu|=500$~GeV and $\lambda=0.1$ we have $|M_{h\hat s}^2|\sim\mathcal O(100\;{\rm GeV})$, which is of order $m_h$ and $m_s$. The situation for smaller $|\mu|$ is not much different.
} 
This is exactly what we wanted to show: for $c_s<1$ the l.h.s.\ of~\eqref{bs_fhs_mix-higgsino} has the sign equal to $-{\rm sgn}(\mu)$
thus the equality cannot hold (and inversely for $c_s>1$). It can be shown (using relations~\eqref{Nj3Nj5}), that the above conclusions hold also when $|\kappa/\lambda|$ is smaller than $|\frac{N_{13}}{N_{15}}\frac{N_{14}}{N_{15}}|$ in eq.~\eqref{bs_fhs_mix} i.e.\ for some part of highly-mixed singlino-Higgsino LSP parameter space. For singlino-dominated LSP we can always choose the sign and value of $\kappa$ to fulfill relation~\eqref{bs_fhs_mix}, however, as we will see below, this may be different in specific versions of NMSSM.

\subsection{Model with $\mu'=0$ \label{model_mup=0}}

Let us now consider the version of NMSSM with $\mu'=0$ in the superpotential~\eqref{W}. It has been shown~\cite{Lee:2011dya} that such superpotential can arise from an underlying $\mathbb{Z}_8^R$ symmetry and hence would be theoretically motivated. Moreover, this model has the same neutralino sector as the no-scale model, while the scalar sector is very similar to the general 
one,\footnote{
Due to the fact that $\mu'$ affects only the diagonal terms in the mass matrix of the CP-even scalars.
} 
which allows us to fix the masses of scalars and simplify greatly our analysis.
\begin{figure}
\center
\includegraphics[width=0.49\textwidth]{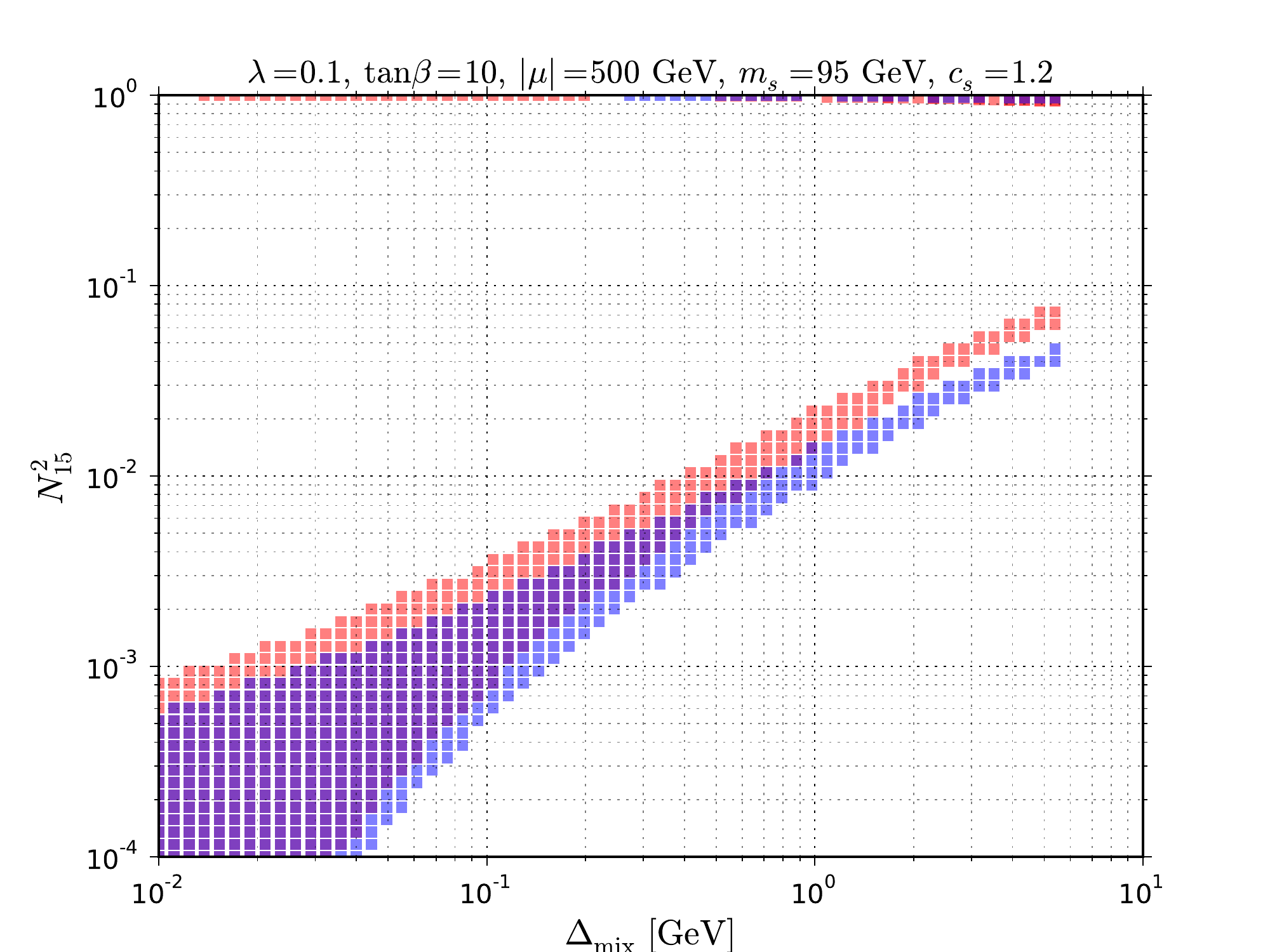}
\includegraphics[width=0.49\textwidth]{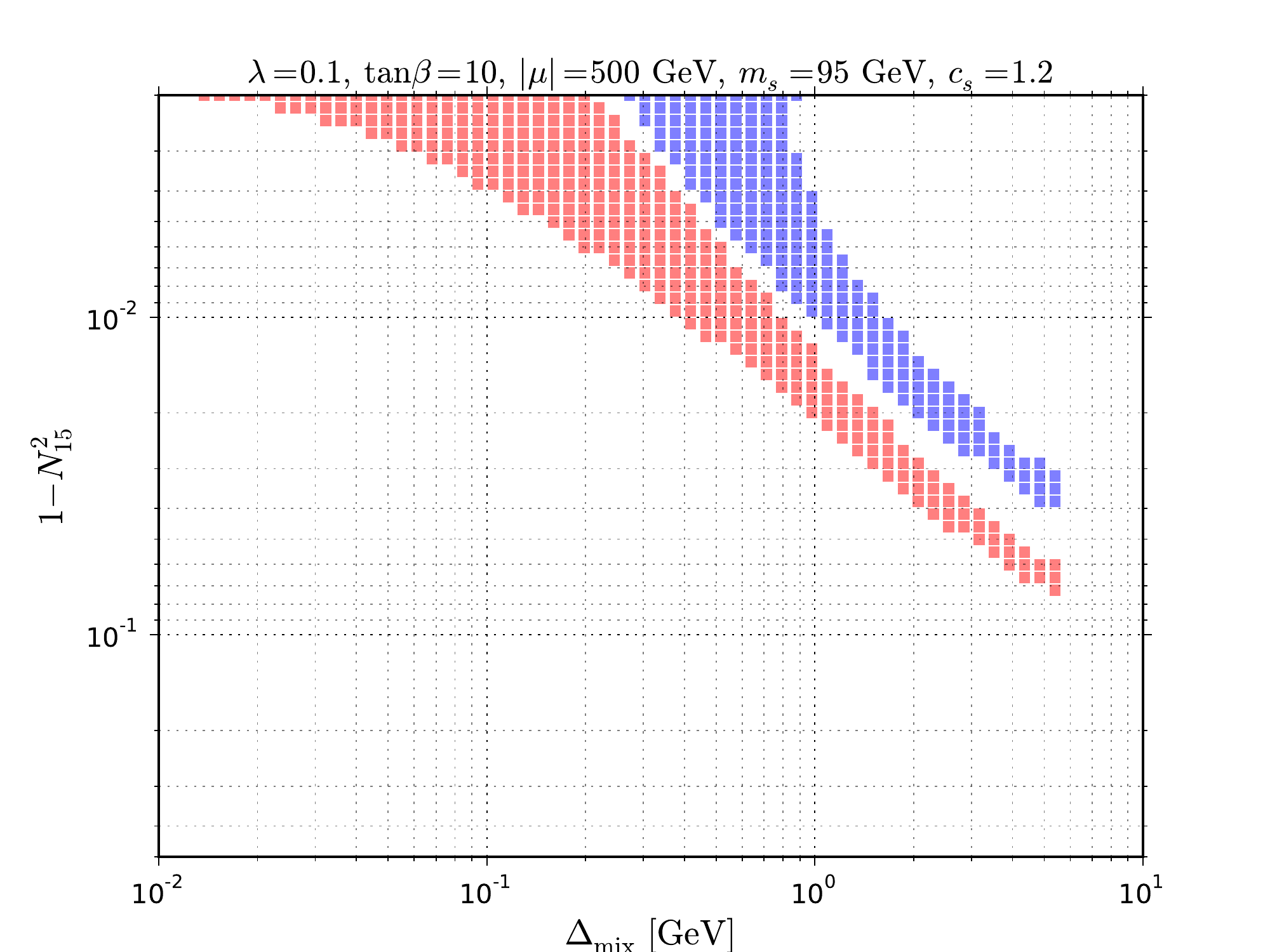}
\caption{Left plot: regions of the plane ($\Delta_{\rm mix}-N_{15}^2$) with $\sigma_{\rm SI}$ smaller than the neutrino background~\cite{NeutrinoB} for $m_{\chi}\mu>0$ (red) and
$m_{\chi}\mu<0$ (blue). Right plot: the same as the left one but vertical axis denotes $1-N_{15}^2$ values i.e.\ the upper, hardly visible region of the left plot is enlarged. We put $\mu'=0$.}
\label{fig:bs_fhs_N15_delta_mup}
\end{figure}

The condition $\mu'=0$ means that the singlino mass parameter is now equal to $\langle\partial_S^2f\rangle=2\mu\kappa/\lambda$. Thus, we can express the ratio $\kappa/\lambda$ in term of $m_\chi/\mu$ in the following way:
\begin{equation}
\label{kaplam}
\frac{\kappa}{\lambda}=
\frac12\left[\left(m_{\chi}/\mu\right)
+\left(\lambda v/\mu\right)^{2}
\frac{\left(m_{\chi}/\mu\right)-\sin(2\beta)}{1-\left(m_{\chi}/\mu\right)^2}
\right]\,.
\end{equation}
Now the $\kappa$ parameter is not an independent one and this may primarily constrain the singlino-dominated LSP region. This is illustrated in Fig.~\ref{fig:bs_fhs_N15_delta_mup} which is the analogue of the right down plot in Fig~\ref{fig:bs_fhs_N15_delta} (we also enlarged the region of singlino-dominated LSP~-- right plot in Fig.~\ref{fig:bs_fhs_N15_delta_mup}). One can see that there is a 
positive correlation between a (small) Higgsino component of the LSP and $\Delta_{\rm mix}$  in the singlino-like LSP region 
(in a general model almost every $\Delta_{\rm mix}$ 
smaller than some value is possible for a highly singlino-dominated LSP).

\begin{figure}
\center
\includegraphics[width=0.49\textwidth]{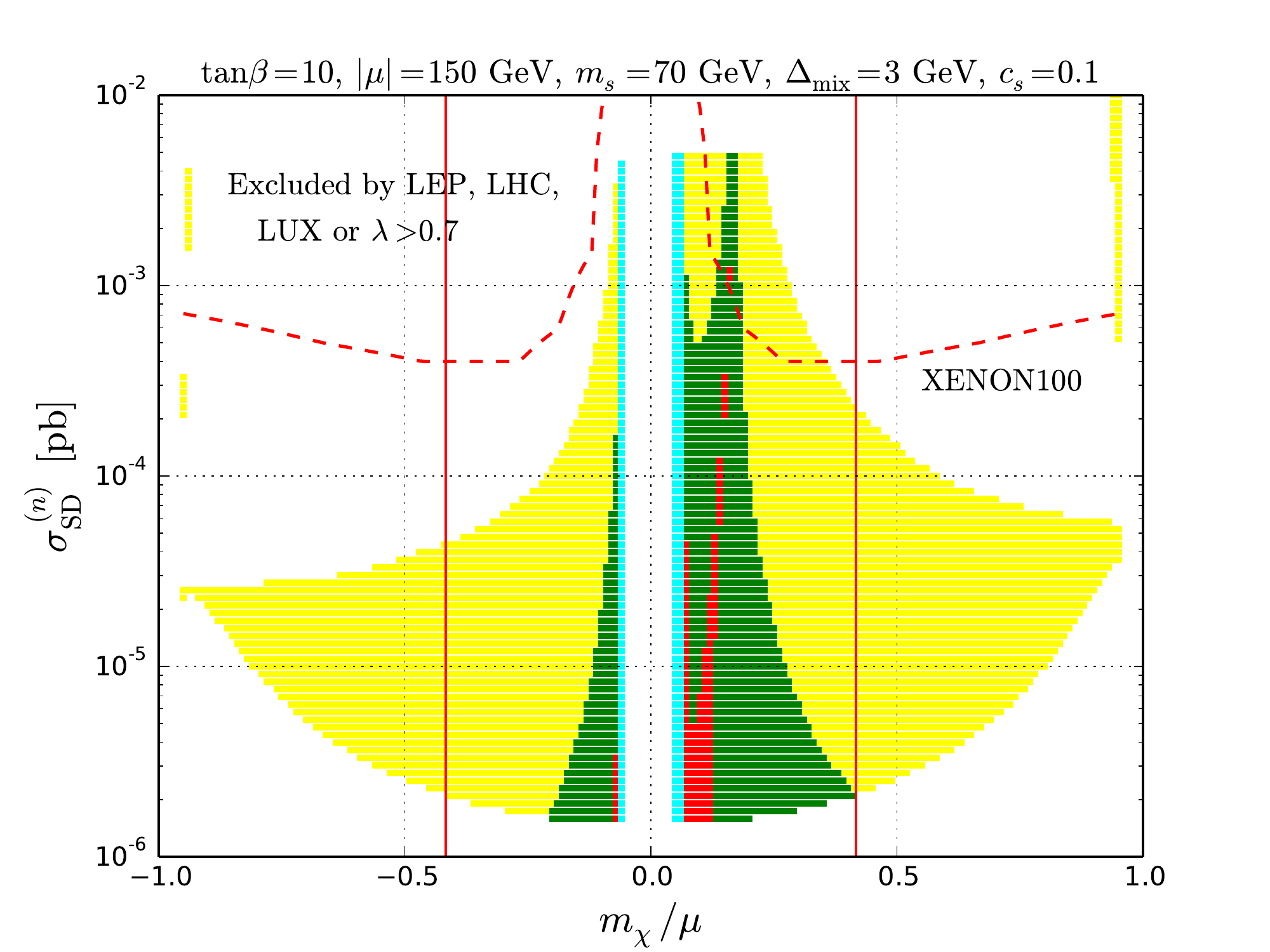}
\includegraphics[width=0.49\textwidth]{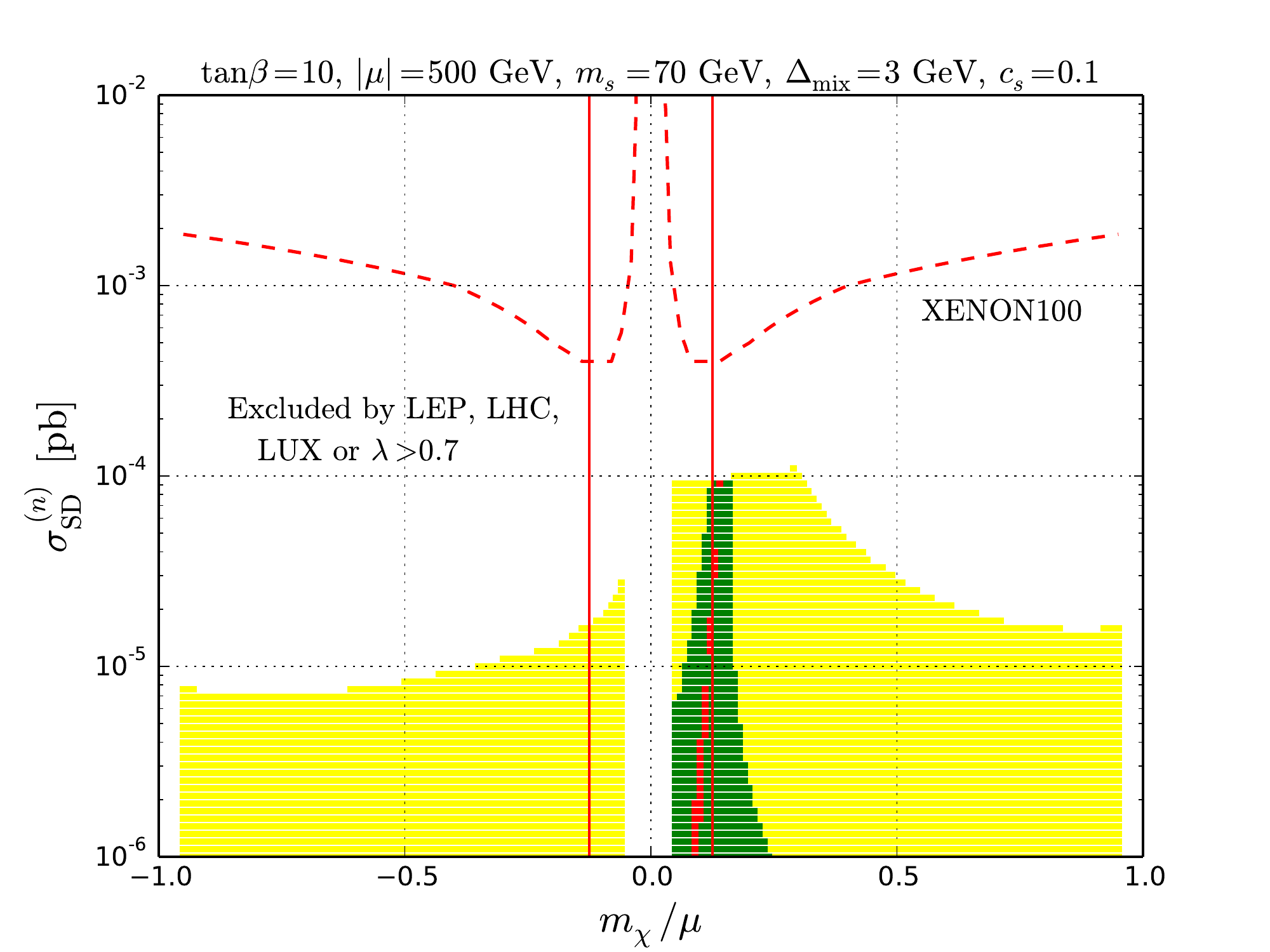}\\
\includegraphics[width=0.49\textwidth]{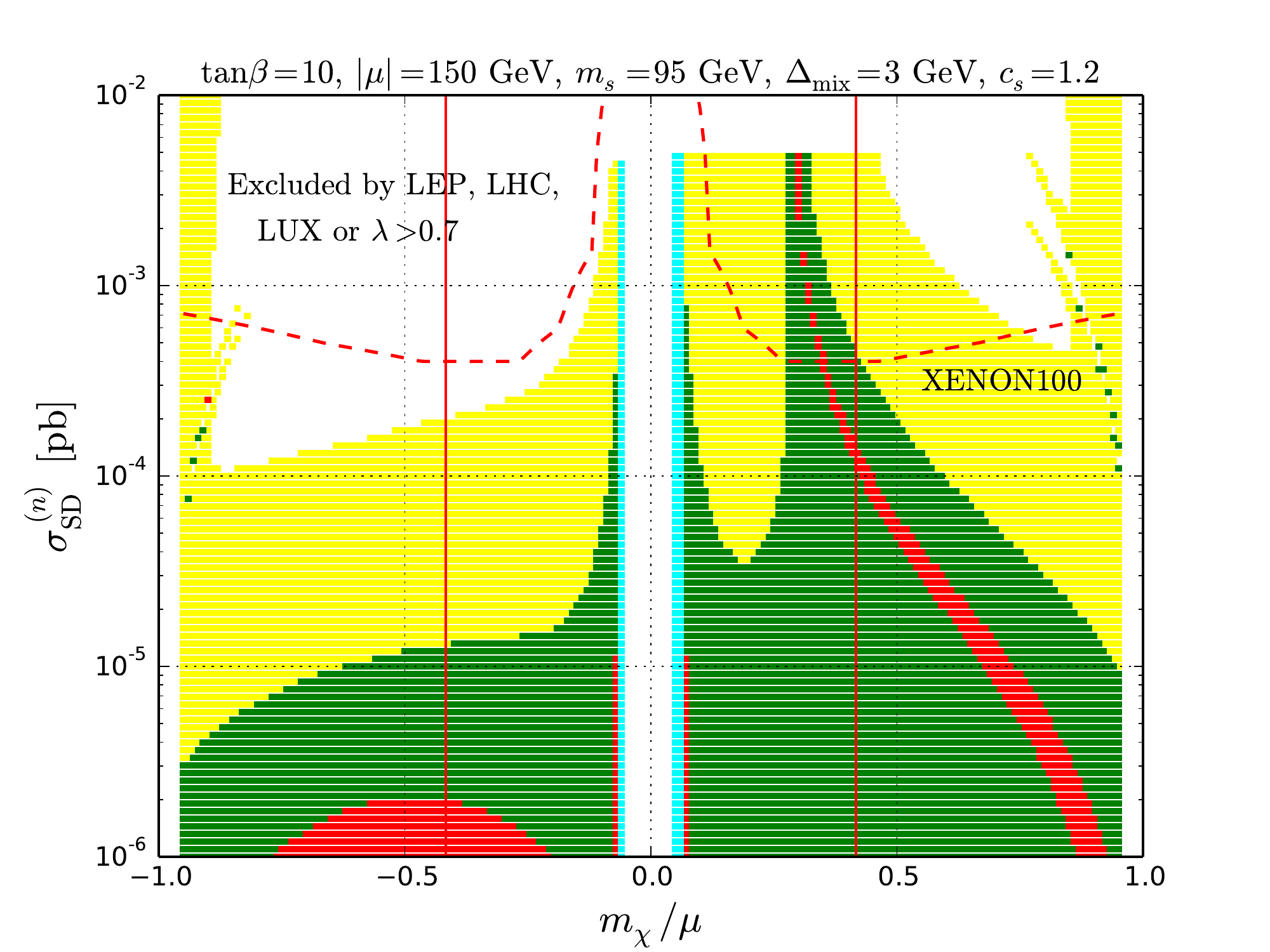}
\includegraphics[width=0.49\textwidth]{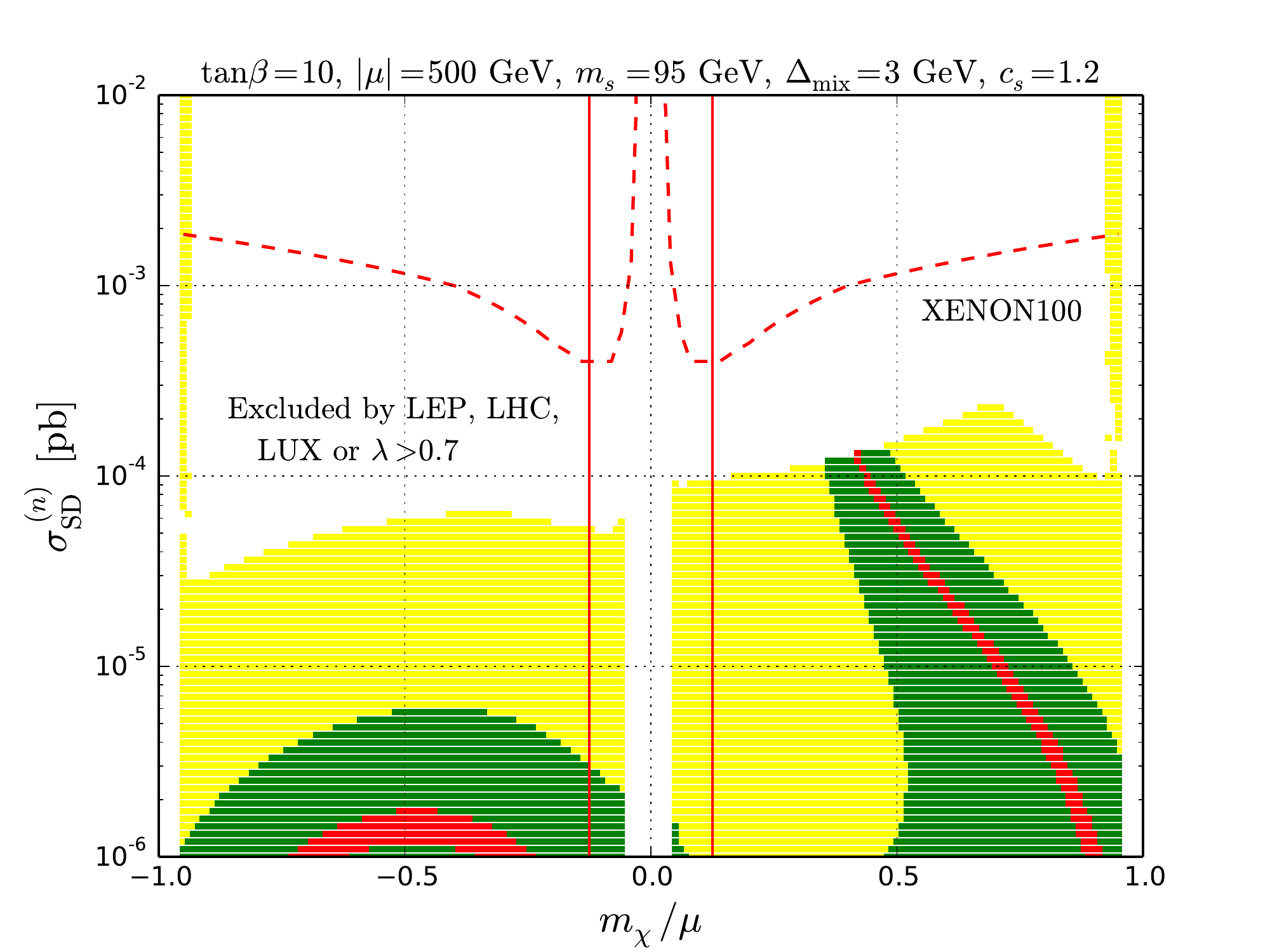}
\caption{Spin-dependent scattering cross section for neutrons $\sigma^{(n)}_{\rm SD}$ versus $m_\chi/\mu$ for the points with $\sigma_{\rm SI}$ below LUX limit~\cite{LUX} (yellow), XENON1T~\cite{XENON1T} limit (green) and neutrino background~\cite{NeutrinoB} (red) for $\mu'=0$. Upper (lower) plots correspond to $m_s=70$ GeV, $c_s=0.1$ ($m_s=95$ GeV, $c_s=1.2$) whereas the left (right) ones to $|\mu|=150$ (500 GeV). Solid red lines represent $m_{\rm LSP}=m_h/2$, whereas dashed ones the XENON100 bound~\cite{XENON100} for $\sigma^{(n)}_{\rm SD}$. Blue regions for $\mu=150$~GeV correspond to light LSP with very weak neutrino background limit $\sim 10^{-46}\;\rm cm^2$.}
\label{fig:bs_fhs_mLSP2mu_SD_mup}
\end{figure}

As we know, for $m_s\lesssim 85$ GeV and sizable $\Delta_{\rm mix}$ we prefer $|c_s|\ll1$ in order to be compatible with the LEP bounds. In a general NMSSM the LSP in this case should be singlino-dominated (see Fig.~\ref{fig:bs_fhs_N15_delta}), where the blind spot condition~\eqref{bs_fhs_mix} can be rather easy fulfilled by tuning the ratio $\kappa/\lambda$. However, for $\mu'=0$ we cannot freely change the value and sign of $\kappa/\lambda$ and therefore fulfill condition~\eqref{bs_fhs_mix} so easily. 
In Fig.~\ref{fig:bs_fhs_mLSP2mu_SD_mup} we present the regions in the plane $(m_\chi/\mu-\sigma^{(n)}_{\rm SD})$ allowing $\sigma_{\rm SI}$ below the LUX bound, XENON1T bound and neutrino background. 
For smaller $\sigma^{(n)}_{\rm SD}$ (and hence smaller $|\lambda v/\mu|$) all the points are singlino-dominated. One can see that for small $|c_s|$ the blind spots are possible only for $m_\chi/\mu$ close to $\sin2\beta\sim2/\tan\beta$ (these are our traditional blind spots in~\eqref{bs_fh_0}), which is not preferable from the 
viewpoint of Higgs invisible decays ($m_{\rm LSP}$ is then smaller than $m_h/2$) unless $|\mu|$ is large: $|\mu|\gtrsim\mathcal{O}\left(500\;\frac{\tan\beta}{10}\right)$~GeV~-- see upper right plot in Fig.~\ref{fig:bs_fhs_mLSP2mu_SD_mup}. We can understand this better if we substitute~\eqref{kaplam} into~\eqref{bs_fhs_mix}, omitting $\frac{N_{13}}{N_{15}}\frac{N_{14}}{N_{15}}$. If $|m_\chi/\mu|$ is small, for large $\tan\beta$ and small $|\lambda v/\mu|$ (which is satisfied in our case) we can neglect the second term in the square bracket in~\eqref{kaplam} and our blind spot condition becomes:
\begin{equation}
\label{bs_fhs_mix-singlino_mup_1}
\frac{m_\chi}{\mu}\left(
1-\frac12\frac{\gamma+\mathcal{A}_s}{1-\gamma\mathcal{A}_s}
\frac{\mu}{\lambda v}
\right)-\sin2\beta\approx0
\,.
\end{equation}
From the sign analysis below~\eqref{bs_fhs_mix-higgsino} it follows that the second term in the bracket in~\eqref{bs_fhs_mix-singlino_mup_1} is positive for $c_s<1$ and hence the smaller $\lambda$ the smaller $|m_\chi/\mu|$ (and reversely for $c_s>1$)~-- one can see this in Fig.~\ref{fig:bs_fhs_mLSP2mu_SD_mup}. On the other hand, if $|m_\chi/\mu|$ is closer to 1 (but LSP is still singlino-dominated), we can neglect the first term in the square bracket in~\eqref{kaplam} and get:
\begin{equation}
\label{bs_fhs_mix-singlino_mup_2}
1\approx
\frac12\frac{\gamma+\mathcal{A}_s}{1-\gamma\mathcal{A}_s}
\frac{\lambda v}{\mu}
\,.
\end{equation}
Note that the above equation does not depend on $m_\chi/\mu$. Again, the sign analysis shows that this kind of a blind spot is possible only for $c_s>1$ (lower plots in Fig.~\ref{fig:bs_fhs_mLSP2mu_SD_mup}). For moderate values of $|m_\chi/\mu|$ both terms in~the square bracket in~\eqref{kaplam} are important and the sign of $m_\chi/\mu$ starts to be crucial e.g.\ for $c_s>1$ and $m_\chi\mu<0$ we need smaller $|\lambda v/\mu|$ to fulfill the corresponding blind spot condition than for $m_\chi\mu>0$. One can see that the XENON100 experiment excludes some region with smaller $|\mu|$  where the SI blind spots are present. The future SD experimental bounds may be especially important for testing the parameter space of the model.

\section{Conclusions}

We have found analytic expressions for blind spots in NMSSM with a light singlet-like scalar and a Higgsino-singlino LSP. Using this formulas, it is relatively
easy to obtain regions with suppressed spin-independent LSP-nucleus cross-section without violating any other experimental constraints, especially when the singlet mass $m_s$ lies
in the LEP favored window i.e.\ in the range of about $85\div100$ GeV (this holds both in a general model and for $\mu'=0$, originating from underlying $\mathbb{Z}_8^R$
symmetry, however in the second case the singlino-dominated region is more constrained). This stays in contrast to the MSSM, where the blind spot conditions put
very severe limitations on the parameter space.

Outside the LEP favored window ($m_s\lesssim 85$ GeV) and when the linear correction to the Higgs mass, $\Delta_{\rm mix}$, is sizable in a general model the LSP tends to be singlino-dominated. In the model with $\mu'=0$ this scenario is additionally constrained by invisible Higgs decays unless $|\mu|$ is large. 

\section*{Acknowledgments}

This work was partially supported by Polish National Science Centre 
under research grants DEC-2012/05/B/ST2/02597, DEC-2014/15/B/ST2/02157
and DEC-2012/04/A/ ST2/00099.


\begin{thebibliography}{99}

\bibitem{BaOlPo}
M.~Badziak, M.~Olechowski and S.~Pokorski,
  JHEP {\bf 1306} (2013) 043
  [arXiv:1304.5437 [hep-ph]].

\bibitem{BS_NMSSM}
  M.~Badziak, M.~Olechowski and P.~Szczerbiak,
  arXiv:1512.02472 [hep-ph].
  
\bibitem{JuKaGr}
  G.~Jungman, M.~Kamionkowski and K.~Griest,
  Phys.\ Rept.\  {\bf 267} (1996) 195
  [hep-ph/9506380].

\bibitem{Belanger:2013oya}
  G.~Belanger, F.~Boudjema, A.~Pukhov and A.~Semenov,
  Comput.\ Phys.\ Commun.\  {\bf 185} (2014) 960
  [arXiv:1305.0237 [hep-ph]].
  
\bibitem{Wagner}
  P.~Huang and C.~E.~M.~Wagner,
  Phys.\ Rev.\ D {\bf 90} (2014) 1,  015018
  [arXiv:1404.0392 [hep-ph]].

\bibitem{Hall}
  C.~Cheung, L.~J.~Hall, D.~Pinner and J.~T.~Ruderman,
  JHEP {\bf 1305} (2013) 100
  [arXiv:1211.4873 [hep-ph]].

\bibitem{NeutrinoB}
  J.~Billard, L.~Strigari and E.~Figueroa-Feliciano,
  Phys.\ Rev.\ D {\bf 89} (2014) 2,  023524
  [arXiv:1307.5458 [hep-ph]].
  
\bibitem{Lee:2011dya}
  H.~M.~Lee, S.~Raby, M.~Ratz, G.~G.~Ross, R.~Schieren, K.~Schmidt-Hoberg and P.~K.~S.~Vaudrevange,
  Nucl.\ Phys.\ B {\bf 850} (2011) 1
  [arXiv:1102.3595 [hep-ph]].

\bibitem{LUX}
  D.~S.~Akerib {\it et al.} [LUX Collaboration],
  Phys.\ Rev.\ Lett.\  {\bf 112} (2014) 091303
  [arXiv:1310.8214 [astro-ph.CO]].
  
\bibitem{XENON1T}
  E.~Aprile [XENON1T Collaboration],
  Springer Proc.\ Phys.\  {\bf 148} (2013) 93
  [arXiv:1206.6288 [astro-ph.IM]].
   
\bibitem{XENON100}
  E.~Aprile {\it et al.} [XENON100 Collaboration],
  Phys.\ Rev.\ Lett.\  {\bf 111} (2013) 2,  021301
  [arXiv:1301.6620 [astro-ph.CO]].
  
\end{thebibliography}
\end{document}